\documentclass[aps,showpacs,pre,twocolumn,reprint,floatfix]{revtex4-1}

\usepackage{amsmath}
\usepackage{graphicx,bm,amssymb,natbib,url,epsfig}
\usepackage{dcolumn}
\usepackage{makecell}

\usepackage{color}
\usepackage{float}

\definecolor{darkgreen}{rgb}{0.1,.6,.1}
\definecolor{greenblue}{rgb}{0.0,.1,.4}
\definecolor{blue-violet}{rgb}{0.54, 0.17, 0.89}

\usepackage[normalem]{ulem}

\allowdisplaybreaks

\begin{document}

\title{Effects of stochasticity and social norms on complex dynamics of fisheries}

\author{Sukanta Sarkar${}^{1}$}
\author{Arzoo Narang${}^{1}$}
\author{Sudipta Kumar Sinha${}^{2}$}\email{sudipta@iitrpr.ac.in}
\author{Partha Sharathi Dutta${}^{1,}$}\thanks{Corresponding author}\email{parthasharathi@iitrpr.ac.in}

\affiliation{ ${}^{1}$Department of Mathematics, Indian Institute of
  Technology Ropar, Punjab, 
  India\\ 
  ${}^{2}$Department of Chemistry, Indian Institute of
  Technology Ropar, Punjab,
  India}%

\received{:to be included by reviewer}
\date{\today}

\begin{abstract}

Recreational fishing is a highly socio-ecological process. Although recreational fisheries are self-regulating and resilient, changing anthropogenic pressure drives these fisheries to overharvest and collapse. Here, we evaluate the effect of demographic and environmental stochasticity for a social-ecological two-species fish model. In the presence of noise, we find that an increase in harvesting rate drives a critical transition from high-yield/low-price fisheries to low-yield/high-price fisheries. To calculate stochastic trajectories for demographic noise, we derive the master equation corresponding to the model and perform Monte-Carlo simulation. Moreover, the analysis of probabilistic potential and mean first-passage time reveals the resilience of alternative steady states. We also describe the efficacy of a few generic indicators in forecasting sudden transitions. Furthermore, we show that incorporating social norms on the model allows moderate fish density to maintain despite higher harvesting rates. Overall, our study highlights the occurrence of critical transitions in a stochastic social-ecological model and suggests ways to mitigate them.

\end{abstract}

\maketitle

\section{Introduction}

Stochasticity, or only noise, is so prevalent that the state of a natural system never remains constant. Noise is often fascinating in its own right and can induce new and unexpected novel phenomena that can not be understood alone from the underlying deterministic skeleton \cite{bjornstad2001noisy,shinbrot2001noise,coulson2004skeletons}. 
Deterministic and stochastic processes interact in exciting ways and share a common origin \cite{black2012stochastic}. Intrinsically noisy and disordered processes generate strikingly regular patterns similar to those systems with ordered processes \cite{shinbrot2001noise}. Further, in the presence of alternative basins of attraction, a stochastic event might shift a system from one attractor to another alternative attractor with contrasting properties \cite{scheffer2001catastrophic, scheffer2009early}. Ecologists are majorly concerned about the temporal fluctuations observed in natural populations, which are the result of intrinsic disturbances (demographic stochasticity) and variations in the global, extrinsic environmental changes (environmental stochasticity)\cite{higgins1997stochastic, myers1998simple, bjornstad1999cycles, lundberg2000population, fromentin2001effects}. Disturbances can modify ecosystems dynamics for various reasons, and predicting ecosystems' response towards these disturbances is a central objective in ecology and conservation. Therefore, understanding the influence of natural and anthropogenic disturbances on ecological variability needs considerable attention. However, failure to elucidate these features of stochastic population dynamics is accountable for the extinction of many vulnerable species, including recreational fisheries.

Increasing anthropogenic pressure on ecosystems has lead to severe biodiversity loss \citep{vitousek1997human,lubchenco1998entering}. Further, harvesting in the absence of adequate fishery management policies has caused overexploitation of many species.  
Although recreational fisheries are, in general, thought to be self-regulating and highly resilient, a rise in overall demand due to human population growth is increasing the vulnerability of these fisheries to sudden collapse \cite{fryxell2017supply}.
According to a recent study \cite{myers1995population}, vast stocks of commercial fisheries in the United States and Europe have witnessed colossal exploitation. Abundance has been severely reduced by harvesting that -- reduced mortality may not be sufficient to allow recovery of a population. Overfishing has driven various species on the verge of extinction, including Atlantic cod, blue whale, threshers, hammerheads \cite{baum2003collapse, gulland1971effect}. Also, excessive harvesting decreases the interaction of overfished populations with other species in the community \cite{cushing1988provident, jackson2001historical}, and eventually drives a sudden decline in their abundance. Some populations might retrieve \cite{clements2019early} but mostly exhibit no signs of recovery, which is indeed a matter of concern.  In the presence of natural stochastic fluctuations, sudden transitions from one equilibrium to another due to crossing a tipping point (i.e., a threshold or a breakpoint) can make a recovery by species challenging. Therefore, analyzing tipping points at which such sudden shifts or critical transitions occur in marine systems is an important area of research \cite{mollmann2012marine, oguz2007abrupt, kraberg2011regime, jiao2009regime, pershing2015evaluating}.

There is now growing awareness about the development of sustainable fishery management policies so that sudden, irreversible decline of fish stocks can be avoided with changing environmental and economic pressures. Since it is well recognized that early warning signals (EWSs) of tipping points in social-ecological systems can advert sudden transitions, they can be used for fishery management \cite{clements2017body,clements2019early}. A lot of efforts has already been made to develop EWSs to forewarn critical transitions in diverse stochastic systems \cite{scheffer2009early,drake2010early,dakos2012methods,clements2017body,sarkar2019anticipating}. Close to a tipping point, the recovery rate from perturbation becomes increasingly slow, termed critical slowing down, resulting in a loss of resilience in a system.  In the presence of small stochastic fluctuations, such critical slowing down is indicated by increasing recovery time, variance, and autocorrelation \cite{scheffer2009early,dakos2012methods}. The effectiveness of EWSs is supported by both theoretical and empirical evidence \cite{dai2012generic,clements2017body,sarkar2019anticipating}. However, after much research on critical transitions and its indicators, the robustness of EWSs is still in question \cite{boettiger2012quantifying,dutta2018robustness}. A thorough understanding of `when' and `how' EWSs work reliably is a major research question.

This article investigates how a socio-ecological prey-predator fish population model responds to harvesting pressure in the presence of stochastic perturbations. Increasing social/human influence on ecological systems is known to exhibit a wide variety of dynamics and critical transitions in their structure and function \cite{lade2013regime}. However, critical transitions and its early warning indicators in coupled socio-ecological systems remain less studied. As the harvesting efficacy pushes a system close to a critical point, humans respond to declining fish-yield by increasing conservationism and making the system stable. This, in turn, allows moderate fish-yield to be maintained despite a higher fishing rate. Mathematically, this can be incorporated in the system by defining social norms and the imitation dynamics of evolutionary game theory \cite{hofbauer1998evolutionary}.

Here, we study the influence of demographic noise into the social-ecological model of fishing \citep{carpenter2017defining} by using the master equation approach and then extending this model for environmental noise. First, we obtain stochastic trajectories for both demographic and environmental noises. Then we determine the resilience of alternative steady states by probabilistic potentials and mean first-passage time. As increasing harvesting rate in the presence of stochastic perturbations drives critical transitions in the model, we examine the efficacy of a few generic early warning indicators using simulated time series. When we simultaneously vary the predation rate and prey harvesting rate, it reveals cyclic dynamics in the system. Further, we also suggest a realizable harvesting strategy by incorporating appropriate social norms into the system for the long-term sustainability of recreational fisheries. Our study shows that proposed social norms can effectively maintain moderate fish-yield despite a high rate of harvesting.

\section{Models and methods \label{s:model}}

\subsection{Deterministic model}

To understand the effects of harvesting on prey and predator fishes, we study a socio-ecological two-species model where both the prey and the predator fishes are exposed to harvesting by fishers. The two-species deterministic model is given as \citep{carpenter2017defining}:
\begin{subequations}\label{Eqn1}
\begin{align}
 \frac{dx}{dt} &= I+ r_x x (1-\frac{x}{K_x})- \frac{p y x^2}{h^2+x^2}-H_x x,\\
 \frac{dy}{dt} &= r_y y (1-\frac{y}{K_y})+\frac{c p y x^2}{h^2+x^2}-H_y y,
\end{align}
\end{subequations}
where $x$ and $y$ represent prey and predator fish densities, respectively. $I$ is the stocking rate, $r_x$ and $r_y$ are the respective growth rates of prey and predator fishes. $K_x$ and $K_y$ are the
maximum potential biomass (or carrying capacity) of $x$ and $y$, respectively. $p$ is predation rate and $h$ is value of $x$ where predation is half maximum. $c$ is the consumption rate of the predator fish. $H_x$ and $H_y$ are harvesting rates of prey and predator fishes, respectively. Understanding dynamics of socio-ecological models is important as humans are responsible for decline in biodiversity, in turn decline in biodiversity badly affects humans.

\subsection{Stochastic model}

While examining temporal changes in population dynamics under natural fluctuations, stochastic modeling comes into the picture. Noise can be introduced in deterministic models in terms of intrinsic (demographic) and/or extrinsic (environmental) factors, and are accountable for causing population fluctuations in the majority of species \citep{may2019stability,lande1993risks}. Demographic stochasticity arises due to an individual's mortality, reproduction, and primarily affects small populations. On the other hand, environmental stochasticity is the variability imposed by the environment on a population and manifested through random fluctuations in population growth rate. Stochasticity in models can bring new, unexpected, and novel phenomena while interacting with deterministic processes. 

\subsubsection{Presence of demographic noise}

To study the influence of demographic noise, we consider different  fundamental processes (e.g., birth and death processes) associated with the model \eqref{Eqn1}, presented in Table~\ref{Table1}. Using all these birth and death processes (gain and loss probability in Table~\ref{Table1}), we develop corresponding master equation for the grand probability function $P(x,y,t)$, and it takes the following form: 
\begin{multline}\label{Eq2}
\frac{\partial P(x, y,t)}{\partial t} = V I  P(x-1, y,t)-V I P(x, y,t) \\+ r_x (x-1) (1-\frac{x-1}{K_x V}) P(x-1,y,t) \\ - r_x x (1-\frac{x}{K_x V}) P(x,y,t) + \frac{p (x+1)^2 y}{(x+1)^2+h^2 V^2} P(x+1,y,t)\\ - \frac{p x^2 y}{x^2+h^2 V^2} P(x, y,t)+ (x+1)H_x P(x+1, y,t)\\ -x H_x P(x, y,t)+ r_y (y-1) (1-\frac{y-1}{K_y V}) P(x, y-1,t) \\- r_y y (1-\frac{y}{K_y V}) P(x, y,t)+ \frac{c p x^2 (y-1)}{x^2+h^2 V^2} P(x, y-1,t)\\- \frac{c p x^2 y}{x^2+h^2 V^2} P(x, y,t)+ (y+1)H_y P(x, y+1,t)\\- y H_y P(x, y,t),
\end{multline}
where $V$ is the volume of the site in which the processes occur.  
Using the van Kampen system's size expansion \citep{van1976expansion}, the master equation \eqref{Eq2} can be written in the following operator notation:
\begin{multline}\label{Eq3}
\frac{\partial P(x,y,t)}{\partial t} = [ V I (E_x^{-1}-1)+ (E_x^{-1}-1) r_x x (1-\frac{x}{K_x V})\\+  (E_x-1) \frac{p x^2 y}{x^2+h^2 V^2}+ H_x (E_x-1) x \\+ (E_{y}^{-1}-1) r_y y (1-\frac{y}{K_y V}) \\ + (E_{y}^{-1}-1) \frac{c p x^2 y}{x^2+h^2 V^2}  +  H_y (E_{y}-1) y ] P(x,y,t),
\end{multline}
where 
\begin{multline*}
E_x P(x,y,t)= P(x+1,y,t)= \rho(\frac{x+1}{V},\frac{y}{V},t)\\= \rho(\hat c_x + \frac{1}{V}, \hat c_y,t) ~~~~~~~~~~~~~\\= \rho(\hat c_x, \hat c_y,t)+ \frac{1}{V} \frac{\partial \rho}{\partial \hat c_x}+ \frac{1}{2 V^2} \frac{\partial^2 \rho}{\partial \hat c_x^2}+... \\ =   \quad(1+ \frac{1}{V} \frac{\partial}{\partial \hat c_x}+
\frac{1}{2 V^2} \frac{\partial^2}{\partial \hat c_x^2}) \rho(\hat c_x, \hat c_y,t) = E_x \rho(\hat c_x, \hat c_y,t)
\end{multline*}
with $E_x= (1+ \frac{1}{V} \frac{\partial}{\partial \hat c_x}+
\frac{1}{2 V^2} \frac{\partial^2}{\partial \hat c_x^2}+ ...)$, $\hat c_x = \frac{x}{V}$, and $\hat c_y = \frac{y}{V}$.
Similarly we can compute $E_x^{-1} \rho(\hat c_x, \hat c_y,t)$, where
$E_x^{-1}= (1- \frac{1}{V} \frac{\partial}{\partial \hat c_x}+ \frac{1}{2 V^2} \frac{\partial^2}{\partial \hat c_x^2}+ ... )$, and $E_y \rho(\hat c_x,\hat c_y,t)$ and $E_y^{-1} \rho(\hat c_x,\hat c_y,t)$.
Using the above expansions, \eqref{Eq3} can be written as:
\begin{multline}\label{Eq4BFP}
\frac{\partial P(x,y,t)}{\partial t} = [\frac{\partial }{\partial
 \hat c_{x}}(-I-\frac{1}{V} r_x x (1-\frac{x}{K_x V}) \\+\frac{1}{V} \frac{p x^2 y}{x^2+h^2 V^2} ~+ \frac{1}{V} H_x x)  \\ + \frac{\partial }{\partial \hat c_{y}}(-\frac{1}{V} r_y y (1-\frac{y}{K_y V})- \frac{1}{V} \frac{c p x^2 y}{x^2+h^2 V^2} + \frac{1}{V} H_y y) \\~+ \frac{1}{2 V^2} \{\frac{\partial^2}{\partial \hat c_{x}^2}(I V + r_x x (1-\frac{x}{K_x V})+ \frac{p x^2 y}{x^2+h^2 V^2}+ H_x x ) \\~+ \frac{\partial^2}{\partial \hat c_{y}^2} ( r_y y (1-\frac{y}{K_y V})+ \frac{c p x^2 y}{x^2+h^2 V^2} +H_y y ) \}  ] P(x,y,t) \;.
\end{multline}

Equation \eqref{Eq4BFP} can also be written in the following compact form as:
\begin{eqnarray*}
\frac{\partial P(x,y,t)}{\partial t} = -\frac{\partial }{\partial
  a}. v(a) P(x,y,t)+ \frac{\partial^{T}}{\partial a}. B. \frac{\partial
}{\partial a} P(x,y,t),
\end{eqnarray*}
which is the Fokker-Planck equation corresponding to the system \eqref{Eqn1}, where
\[ v(a)=
\left[ {\begin{array}{c}
      I+\frac{1}{V} r_x x (1-\frac{x}{K_x V}) -\frac{1}{V} \frac{p x^2 y}{x^2+h^2 V^2}- \frac{1}{V} H_x x \\
       \frac{1}{V} r_y y (1-\frac{y}{K_y V})+ \frac{1}{V} \frac{c p x^2 y}{x^2+h^2 V^2} - \frac{1}{V} H_y y \\
        \end{array} } \right], \]
\[ \frac{\partial }{\partial a}=
\left[ {\begin{array}{c}
      \frac{\partial }{\partial \hat c_{x}} \\
      \frac{\partial }{\partial \hat c_y} \\
\end{array} } \right], \]
and
\[B= \left[ {\begin{array}{cc}
      b_{11}  & 0 \\
      0 &  b_{22} \\
\end{array} } \right]\]
with $ b_{11} =  I V + r_x x (1-\frac{x}{K_x V})+ \frac{p x^2 y}{x^2+h^2 V^2}+ H_x x$ and $b_{22} =  r_y y (1-\frac{y}{K_y V})+ \frac{c p x^2 y}{x^2+h^2 V^2} +H_y y$.

{\renewcommand{\arraystretch}{1.5}
\begin{table*}
\centering
\caption{Different birth and death processes, change of state vectors, gain and loss probabilities, and their propensity function, associated with the prey-predator fish model \eqref{Eqn1}.  $V$~is the system's volume where all the reactions occur. The symbols ($+1$)~and ($-1$)~in the column of state vectors represent birth and death processes of the respective chemical species. Here, $P$ stands for the grand probability function.  $\phi$ and $z$ are empty state and dummy variables, respectively. }\label{Table1}
\begin{center}
\begin{tabular}{| m{0.5cm} | m{2.1cm} | m{2cm} | m{1.4cm} | m{1.4cm} | m{3.28cm} | m{2.5cm} | m{2cm}|}
\hline Sl. No. & \centering Elementary events & Reaction type & \makecell{Before \\ reaction} & \makecell{After\\ reaction} & \centering Gain probability & \centering Loss probability & \makecell{Propensity\\ function $(a_n)$}\\

\hline ~1. &  \centering $\phi \rightarrow x$ &\makecell{Growth of \\prey \\ $~~~$} &  \centering $\begin{bmatrix} x-1\\ y \end{bmatrix}$  &  \centering $\begin{bmatrix} x\\ y \end{bmatrix}$  &  \centering $ V I P(x-1,y)$ &  \centering $V I P(x, y)$ &   $~~~~~~ V I$ \\

\hline ~2. & $x + \phi \rightleftharpoons$ $x+x$ & \makecell{Logistic\\ growth\\of prey} &  \centering $\begin{bmatrix} x-1\\ y \end{bmatrix}$  &  \centering $\begin{bmatrix} x\\ y \end{bmatrix}$ &  \centering $\makecell{r_x (x-1) (1-\frac{x-1}{K_x V}) \\ \times P(x-1,y)}$ &  \centering $\makecell{r_x x (1-\frac{x}{K_x V})\\ \times P(x,y)}$ & $r_x x (1-\frac{x}{k_x V})$  \\

\hline ~3. & \makecell{$2 x+y \rightarrow z$ \\$z \rightarrow x$} & \makecell{Holling\\ type III\\response for\\ predation} &  \centering $\begin{bmatrix} x+1 \\ y \end{bmatrix}$  &  \centering $\begin{bmatrix} x \\ y \end{bmatrix}$ &$ \frac{p (x+1)^2 y}{(x+1)^2+h^2 V^2}  P(x+1,y)$ &  \centering $\frac{p x^2 y}{x^2+h^2 V^2}P(x,y) $ & ~~~ $\frac{p x^2 y}{x^2+h^2 V^2} $ \\

\hline ~4. &  \centering $x \rightarrow \phi$  & \makecell{Mortality of\\ prey due to \\harvesting} &  \centering $\begin{bmatrix} x+1\\ y \end{bmatrix}$  &   \centering $\begin{bmatrix} x\\ y \end{bmatrix}$  &  \centering $(x+1) H_x P(x+1,y) $ &  \centering $x H_x  P(x,y) $ & $ ~~~~~~ x H_x $ \\

\hline ~5. & $y + \phi \rightleftharpoons$ $y+y$ &\makecell{Logistic\\ growth\\of predator} &  \centering $\begin{bmatrix} x\\ y-1 \end{bmatrix}$  &  \centering $\begin{bmatrix} x\\y \end{bmatrix}$  &  \centering $\makecell{r_y (y-1) (1-\frac{y-1}{K_y V}) \\ \times P(x, y-1)}$  &  \centering $\makecell{r_y y (1-\frac{y}{K_y V}) \\ \times P(x,y)}$& $ r_y y (1-\frac{y}{K_y V})$   \\

\hline ~6. & \makecell{$2 x+y \rightarrow z$ \\$z \rightarrow y+y$} & \makecell{ Holling\\ Type III\\ process of\\ predator} &  \centering $\begin{bmatrix} x\\ y-1 \end{bmatrix}$   &  \centering $\begin{bmatrix} x \\ y \end{bmatrix}$  &  \centering $\frac{c p x^2 (y-1)}{x^2+h^2 V^2} P(x,y-1)$ &  $\frac{c p x^2 y}{x^2+h^2 V^2} P(x,y)$  &  $~~~~ \frac{c p x^2 y}{x^2+h^2 V^2}$ \\

\hline ~7. &  \centering $y \rightarrow \phi$  & \makecell{Mortality of\\ predator due\\ to Harvesting} &   \centering $\begin{bmatrix} x\\ y+1 \end{bmatrix}$  &   \centering $\begin{bmatrix} x\\ y \end{bmatrix}$  &  \centering $(y+1)H_yP(x,y+1)$  &  \centering $y H_yP(x,y) $ & $~~~~~~y H_y $ \\
\hline
\end{tabular}
\end{center}
\end{table*}
}

\subsubsection{Presence of environmental noise}
 
Ecosystems are often influenced by environmental factors such as, climate change, weather conditions, etc.  Stochastic fluctuations in population due to such external variations are termed as environmental noise, and it is known to be one of the major sources affecting the resilience of a system \cite{boettiger2018noise}.  To study the influence of environmental noise we formulate a stochastic model of \eqref{Eqn1} with multiplicative noise. The stochastic model is given as:
\begin{equation}\label{EqEN}
\frac{dX}{dt}= F(X) + \sigma X \xi_t\;,
\end{equation}
where $X=[x,y]^{T}$ represents prey-predator fish vector. $F(X)$ is the
interaction between prey and predator fish given by the right hand side of
\eqref{Eqn1}.  $\sigma = [\sigma_x,\sigma_y]^T$ is noise intensity, and $\xi_t$ is the Gaussian white
noise with zero mean and unit variance. The Fokker-Planck equation corresponding to \eqref{EqEN} can be written in a straightforward manner. The final expression of the Fokker-Planck equation takes the following form:
\begin{eqnarray*}
\frac{\partial P(x,y,t)}{\partial t} = -\frac{\partial }{\partial
  a}. v(a) P(x,y,t)+ \frac{\partial^{T}}{\partial a}. B. \frac{\partial
}{\partial a} P(x,y,t),
\end{eqnarray*}
where
\[ v(a)=
\left[ {\begin{array}{c}
      I+\frac{1}{V} r_x x (1-\frac{x}{K_x V}) -\frac{1}{V} \frac{p x^2 y}{x^2+h^2 V^2}- \frac{1}{V} H_x x \\
       \frac{1}{V} r_y y (1-\frac{y}{K_y V})+ \frac{1}{V} \frac{c p x^2 y}{x^2+h^2 V^2} - \frac{1}{V} H_y y \\
        \end{array} } \right], \]
\[ \frac{\partial }{\partial a}=
\left[ {\begin{array}{c}
      \frac{\partial }{\partial \hat c_{x}} \\
      \frac{\partial }{\partial \hat c_y} \\
\end{array} } \right], \]
and
\[B= \left[ {\begin{array}{cc}
     \hat b_{11}  & 0 \\
      0 &  \hat b_{22} \\
\end{array} } \right],\]
with $\hat b_{11} =  \sigma_x^2 x^2$ and $\hat b_{22} = \sigma_y^2 y^2$.

\subsection{Stochastic simulation methods}
It is a well known fact that presence of noise in a bistable system can trigger a tipping \cite{scheffer2001catastrophic}, which will be evident from its trajectories. In the presence of demographic noise, stochastic trajectories can be generated by solving the master equation \eqref{Eq2}. We have used the Monte-Carlo simulation \citep{gillespie1976general,van1992stochastic} to numerically solve the master equation \eqref{Eq2}, and finally generate stochastic trajectories. The simulation was performed using the Gillespie algorithm \cite{gillespie1976general}. The algorithm works in the following way: it first calculate the propensity functions ($a_n$) in Table~\ref{Table1} and then draw two uniform random numbers ($r_{1}$ and $r_{2}$) from the interval $(0,1)$. One random number, say $r_{1}$ will be used to update the time $(t)$ for the next process stochastically, where 
 $ \tau=\frac{1}{a_0}\ln(\frac{1}{r_1})$, and $a_0=\sum_{j=1}^{n} a_j$. The other number $r_{2}$ is used to identify the index $n$~of the occurring process and which is given by the smallest integer satisfying:
\[
\sum_{j=1}^{n -1} a_j <r_2a_0 \leq \sum_{j=1}^{n} a_j\;.  
\]
The system states are updated $X(t+\tau)=X(t)+\Delta X_{\nu_n}$, then the simulation proceeds to the next occurring time. $\Delta X_{\nu_n}$~represents change in species vector in the stochastic time interval $\tau$.

To calculate trajectories of the model in the presence of environmental noise, we performed stochastic simulations of \eqref{EqEN} in MATLAB (R2018b) using the Euler-Maruyama method with an integration step size of 0.01.

\subsection{Probabilistic potentials}

Here, we try to understand the influence of demographic noise by calculating probabilistic potentials for different prey harvesting rates ($H_x$). Probabilistic potentials are calculated from numerically simulated trajectories. These trajectories are computed for $4 \times 10^4$ different initial conditions, which are uniformly drawn from [$0$, $6$] $\times$ [$1.5$, $2.3$]. The probability is then calculated by counting the number of visits of a trajectory to a steady state for all initial conditions. In particular, if $x^{ss}$ and $y^{ss}$ are respective steady state densities of prey and predator fishes, $M^{ss}_{xy}$ is the frequency count to visit a steady state ($x^{ss},y^{ss}$) and $N$ is the total number of visits. Then probability is defined as $P_{ss}(x^{ss},y^{ss})=\frac{M^{ss}_{xy}}{N}$. The negative logarithm of this probability is the desired stochastic potential.

\subsection{Mean first-passage time}

In a bistable region or a bistable potential, two steady states are separated by an unstable steady state defining the barrier state or the potential barrier.  In the presence of noise, often exit from a potential well is possible. In this regard, one can calculate the first passage time (FPT).  The FPT is calculated by considering a family of initial points, ($x_0, y_0$) that starts somewhere in a potential well of volume $V$ in space, bounded by a surface $\partial V$. Then the FPT is the time when the point exits $V$ for the first time. The mean first-passage time (MFPT) is the average time of the FPT.  The permanence of steady states in a noise-driven system is often characterized by the MFPT \citep{gardiner1985handbook}.  Ideally, the motion of the family of initial points satisfies both of the master and Fokker-Planck equations and we focus only those points that have not left $V$~by time $t$. We remove all the paths that have crossed the boundary of $V$~before time $t$, which is usually done by placing an absorbing boundary condition on $\partial V$. We impose the absorbing wall at the potential
barrier of the bistable system. We perform Gillespie simulation for a family of
initial points, which allows us to find the time taken to cross the absorbing wall. Here, the mean time is calculated over $5 \times 10^3$ realizations, and all the initial conditions are taken uniformly from the ranges $[0,6] \times [1.5,2.3]$.
 
\subsection{Early warning signals}

The presence of noise in a bistable system can drive tipping, and it can be predicted by critical slowing down based EWSs. We calculate a few generic EWSs, like variance, lag-1 autocorrelation (AR(1)), and conditional heteroskedasticity to forewarn upcoming transitions. Variance $(\sigma^2)$ is the variability of a random variable $x$ around the mean ($\mu$) and is given by $ 
\sigma^2=\frac{1}{N} \sum_{n=1}^{N}[x(t)- \mu]^2 $,
 where $N$ is number of time step. We also calculate the autocorrelation at lag-$1$, which is given by
$ \rho=\frac{E \{[x(t+1)-\mu][x(t)- \mu]\}}{\sigma^2}$,
where $E$ is expectation value of a random variable, $x(t)$ is the value of the state variable at time $t$. To get robust predictions, we also calculate conditional heteroskedasticity following methods in \citep{seekell2011conditional}. It uses the Lagrange multiplier test to calculate the conditional heteroskedasticity.

\begin{figure}
\begin{center}
\includegraphics[width=1.1\columnwidth]{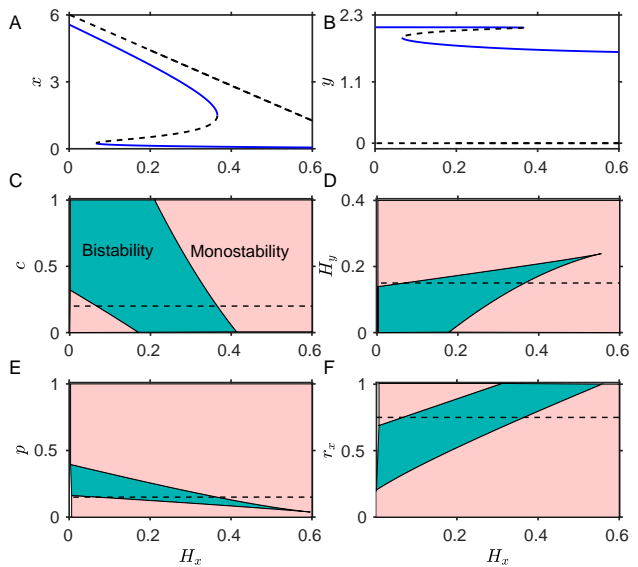}
\caption{\label{Fig1}  Bifurcation diagrams of the model \eqref{Eqn1}. Co-dimension one bifurcation diagram of (A) the prey fish ($x$), and (B) the predator fish ($y$), for changes in $H_x$.  (C-F) Co-dimension two bifurcation diagrams in the following $c-H_x, ~ H_y-H_x ~ p-H_x$ and $r_x- H_x$ planes, respectively. In (A) and (B), solid and dashed curves denote stable and unstable states. In (C- F), the darker regions denote bistability, and other regions denote monostability. Dashed lines in two-parameter bifurcation diagrams denote the values used for numerical simulation.}
\end{center}
\end{figure}

\begin{figure*}[!ht]
\begin{center}
\includegraphics[width=0.9\textwidth]{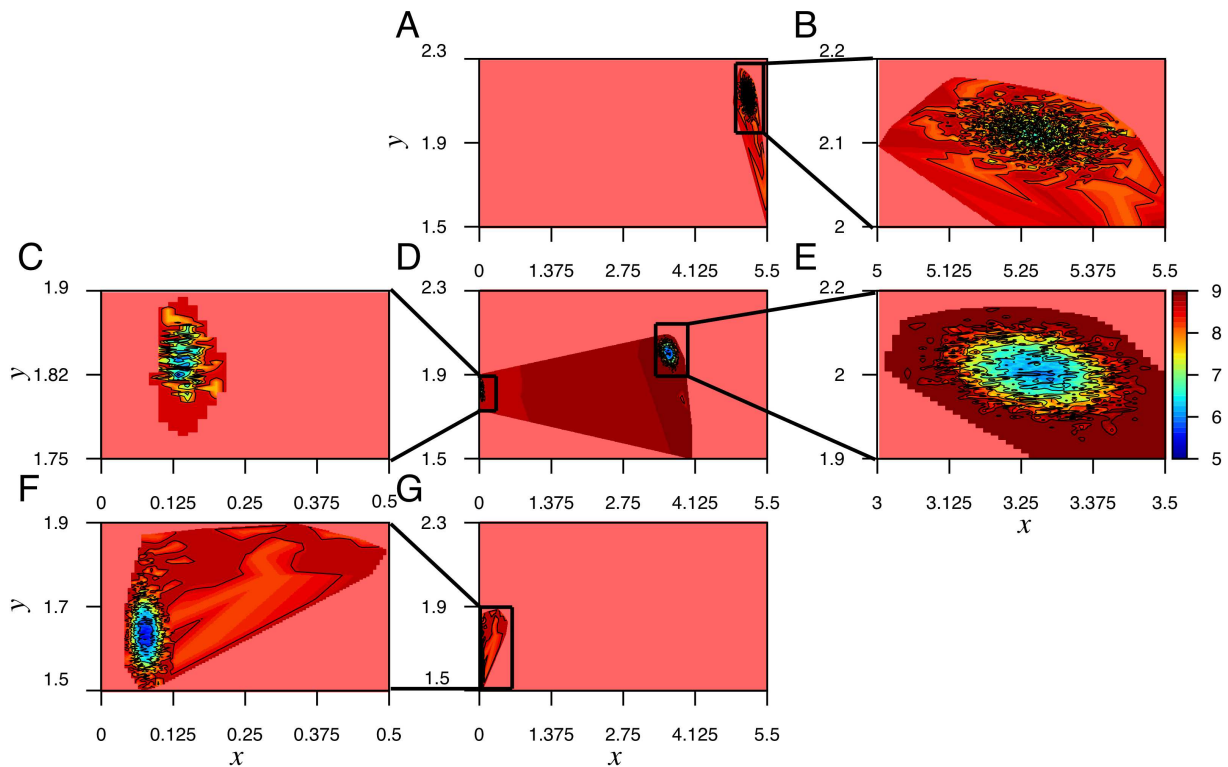}
\caption{\label{Fig2} Stochastic potential landscapes for different values of the prey harvesting rate ($H_x$), obtained by solving the master equation \eqref{Eq2}. Stochastic potential for: (A) monostable high density fish state at $H_x=0.03$, (D) bistable high density and low density fish states for $H_x=0.2$, and (G) monostable low density fish state for $H_x=0.45$. The blowup diagrams (B, C, E, F) magnify regions in potential wells. }
\end{center}
\end{figure*}

\begin{figure}[!ht ]
\begin{center}
\includegraphics[width=0.8\columnwidth]{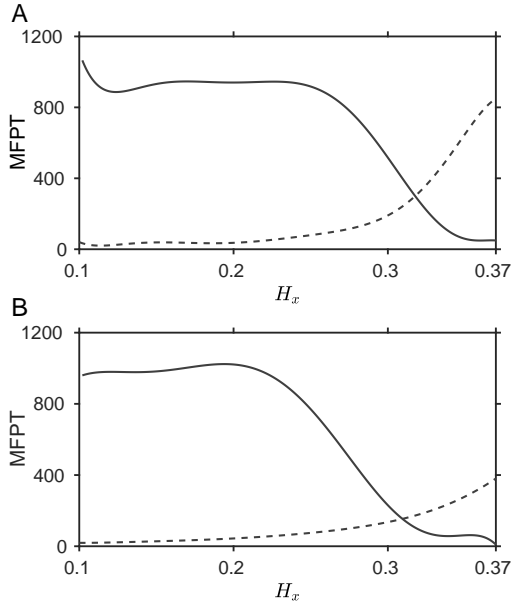}
\caption{\label{Fig3}  Mean first-passage time with an increase in $H_x$: (A) for demographic noise, and (B) for environmental noise.  Solid curve shows the MFPT taken by the system to switch from the upper steady state to the lower steady state.  Similarly dotted curve shows MFPT taken by the system to switch from the lower steady state to the upper state. Environmental noise intensity is taken as $\sigma =0.03$ }
\end{center}
\end{figure}
\begin{figure}[!ht]
\begin{center}
\includegraphics[width=0.8\columnwidth]{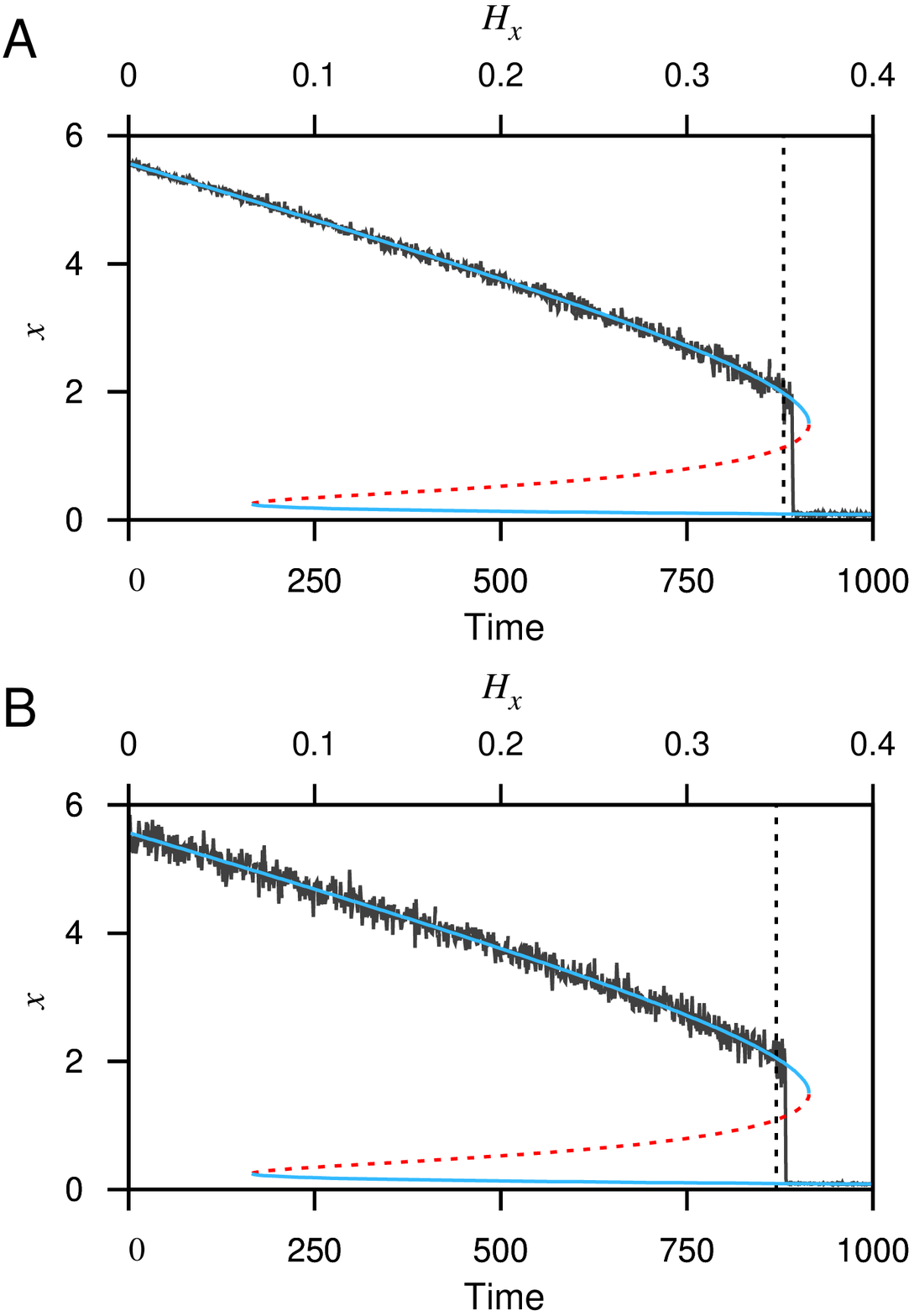}
\caption{\label{Fig4} Critical transitions in the presence of: (A) demographic noise, and (B) environmental noise, with increasing harvesting rate $H_x$. Cyan solid lines and red dashed lines represent stable and unstable steady states, respectively.  Black dashed vertical line represents the position upto where the times series data has taken for statistical analysis.  Noise intensity for the environmental noise is taken as $\sigma =0.03$.}
\end{center}
\end{figure}

 \section{Results \label{res}}
 
Depending on the parameter values and the Holling Type-III functional response, the model \eqref{Eqn1} exhibits monostability as well as bistability (see Fig.~\ref{Fig1}). The system evolves from a high-density state to a low-density state of fish via a hysteresis loop with an increase in the harvesting rate. With a low to moderate harvest, there is a high positive equilibrium. With overharvest, the high positive equilibrium suddenly collapses into a low positive equilibrium. Unless otherwise stated, values of the parameters used throughout this study are: $I=0.01$, $r_x=0.75$, $K_x=6$, $h=0.15$, $r_y=0.25$, $K_y=4$, $c=0.2$, $p=0.15$, and $H_y=0.15$.

 \subsection{Stochastic potentials}
 
Stochastic potentials of the system \eqref{Eqn1} for different $H_x$ are depicted in Fig.~\ref{Fig2}. Potentials show a clear effect of stochasticity on the model's behavior. Increasing the harvesting rate changes the probabilistic potential, each for low and high values of $H_x$; there exists one potential well. For an intermediate value of $H_x$, there exist two potential wells. The lowest value depicted in the potential color-bar corresponds to the existence of a deep well, i.e., the likelihood of a return to the steady state, after perturbation, is more in this well. It is evident that, in the monostable regions, the high-density state's well is flatter and shallower (Figs.~\ref{Fig2}A and \ref{Fig2}B), and in comparison to that the low-density well is deeper (Figs.~\ref{Fig2}F and \ref{Fig2}G). However, in the bistable region, the potential well for the high-density state is deeper than the low-density state (Figs.~\ref{Fig2}C, \ref{Fig2}D and \ref{Fig2}G). This deeper well represents higher resilience and indicates that a population near the high-density state is more likely to stay there unless it is largely perturbed in the bistable region.

\subsection{Mean first-passage time}

The effects of changing $H_x$ on MFPT under intrinsic and extrinsic stochastic
fluctuations are shown in Figs.~\ref{Fig3}A and \ref{Fig3}B. The reciprocal of MFPT is the rate of arrival at the potential barrier. The mean time taken for both the noises to shift from upper to lower steady state decreases with an increase in the harvesting rate of prey, whereas the mean time decreases with decreasing harvesting rate to traverse
from a lower to an upper stable state. In other words, the rate of transition from
the upper steady state to the lower steady state increases with an increase in the prey's harvesting rate. However, the transition from lower to upper steady state happens quite rarely since the rate is slow. It is also evident that MFPT for the high-density state is much higher than the low-density state, indicating higher resilience of the high-density state. The crossing point of MFPTs for two different noises are different. In the case of environmental noise, the point is earlier in comparison to demographic noise (Figs.~\ref{Fig3}A and ~\ref{Fig3}B). This establishes that MFPT changes with noise type and the fluctuations observed under environmental stochasticity are more than that of demographic stochasticity.

\begin{figure*}[]
\begin{center}
\includegraphics[width=1.95\columnwidth]{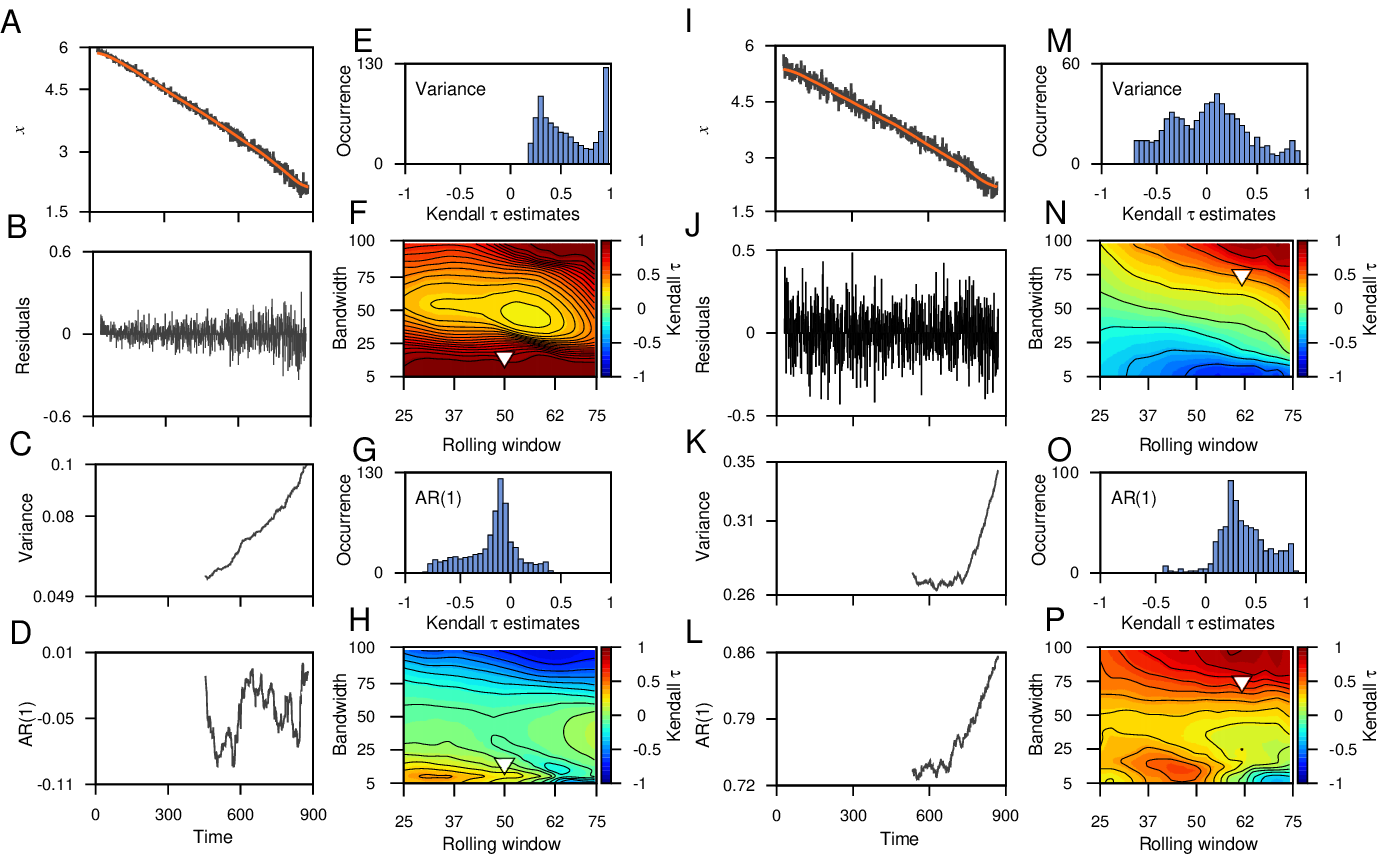}
\caption{\label{Fig5} Early warning signals were calculated from the time series data for demographic and environmental noises. (A, I) Simulated time series data before a transition. (B, J) Residual time series after applying a Gaussian filter (red curve in (A, I) is the trend used for filtering). (C, D) Variance and AR(1) are calculated using a moving window size of $50\%$ and bandwidth $10$. Similarly, (K, L) represents variance, and AR(1) calculated using a moving window size $60\%$ and bandwidth 75. (E, G, M, O) Distribution of the Kendall-$\tau$ estimates. (F, H, N, P) Sensitivity analysis for the combination of rolling window size and bandwidth.}
\end{center}
\end{figure*}

\begin{figure}[h!]
\begin{center}
\includegraphics[width=0.98\columnwidth]{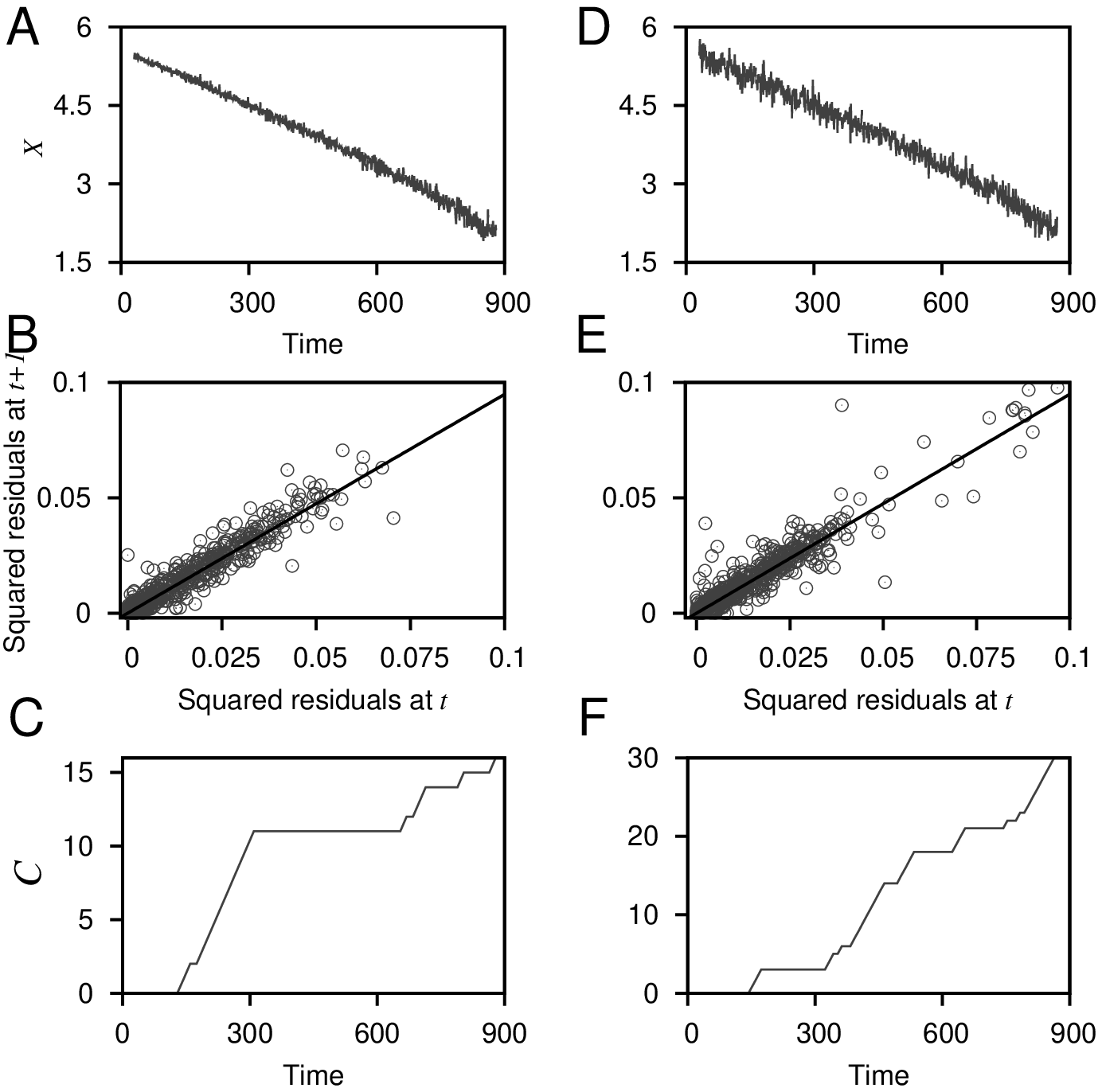}
\caption{\label{Fig6} Time series before critical transitions for (A) demographic noise, and (D) environmental noise. (B, E) Correlation between squared residuals at time step $t$ with the next time step ($t+1$) from an autoregressive lag-$1$ model and slanted lines present the linear regression lines. (C, F) Cumulative Lagrange multiplier test ($C$) for conditional heteroskedasticity.}
\end{center}
\end{figure}

\subsection{Early warning signals}

Here we explore the efficacy of EWSs in anticipating regime shifts by analyzing stochastic time series data (see Fig.~\ref{Fig4}) for both the demographic and environmental noise. To calculate variance and AR($1$), we have selected a time series segment ( Figs.~\ref{Fig4}A and \ref{Fig4}B) before a transition. Indicators of an impending transition are affected by non-stationarities in the mean of the time series, especially the metrics calculated in a moving windows \citep{dakos2012methods}. So, it is always appropriate to remove non-stationarities as it imposes a formidable correlation structure on time series. Hence, for our analysis, all metrics calculated within rolling windows have removed high frequencies using the Gaussian detrending with a bandwidth. Then we calculate variance (see Figs.~\ref{Fig5}C and \ref{Fig5}K) and AR($1$) (see Figs.~\ref{Fig5}D and \ref{Fig5}L) using rolling windows of $50\%$ and $60\%$ of the size of time series, respectively.

We see that under both kinds of stochasticity (demographic and environmental), the variance is robust and successfully estimates an impending transition (see Figs.~\ref{Fig5}C and \ref{Fig5}K). However, we find that AR($1$) is only successful in predicting regime shifts in the case of environmental stochasticity as depicted in Fig.~\ref{Fig5}L, but does not work well for demographic noise (see Fig.~\ref{Fig5}D). This asserts that the commonly used CSD indicators are not always robust in predicting an upcoming transition.

Therefore we use another indicator called conditional heteroskedasticity, which minimizes false-positive signals from the time series. It is the conditional error variance in time series, i.e., error variance at a time step is dependent or conditional on the error variance of the previous time step. We consider the time series data for demographic and environmental noise (see Figs.~\ref{Fig6}A and \ref{Fig6}D) to calculate conditional heteroskedasticity. We took a rolling window of size $10$ of the time series while calculating the conditional heteroskedasticity. To remove stationarity from the data set, we apply the Gaussian kernel fitting to the time series and calculate its residual square. Here squared residuals at time step $t$ is plotted with the previous time step (see Figs.~\ref{Fig6}B and \ref{Fig6}E). The positive correlation between these two squared residuals suggests that both time series are having conditional heteroskedasticity. Hence, the cumulative number of significant Lagrange multiplier tests ($C$) is carried out on both data sets, as shown in Figs.~\ref{Fig6}C and \ref{Fig6}F. We observe an increasing trend in $C$ for both the time series, thus indicating that both the time series are conditionally heteroskedastic and approach a critical transition.

\subsection{Collective influence of predation and harvesting on the dynamics of fisheries}

\begin{figure}
\begin{center}
\includegraphics[width=1\columnwidth]{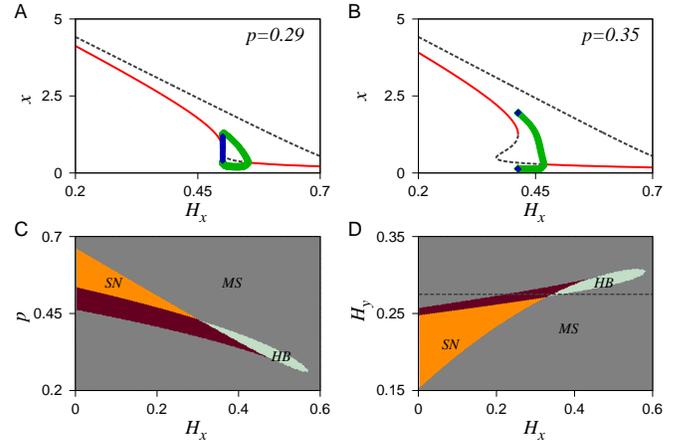}
\caption{\label{fig7} Codimension-one bifurcation diagrams of the prey fish $x$ with variations in $H_x$, for different values of $p$: (A) $p=0.29$, and (B) $p=0.35$, with $H_y=0.275$ (along the dashed line shown in subfigure (D)). (C-D) Codimension-two bifurcation diagrams in the $H_x - p$ plane and $H_x - H_y$ plane, respectively. In (A-B), red solid curves denote stable steady state, and black dashed curves denote unstable steady state. Green and blue circles are the stable and unstable limit cycles, respectively. In (C-D), MS, SN, and HB represent monostable, saddle-node bifurcation, and Hopf bifurcation curves, respectively. }
\end{center}
\end{figure}

Fish stocks are greatly influenced by two major factors -- predation and harvesting. Therefore, illustrating the criticality of these two processes is significant in regulating fish populations. Here we explore the qualitative dynamics of the two species model \eqref{Eqn1} with varying the predation rate $p$ and the harvesting rate $H_x$ (see Fig.~\ref{fig7}). As can be seen from Figs.~\ref{fig7}A and \ref{fig7}B, an increase in $H_x$ eventually leads to the collapse of prey fish. However, the occurrence of fluctuations in prey densities is governed by the predation level. In Fig~\ref{fig7}A, for $p = 0.29$, the oscillations are observed around $H_x = 0.47$, whereas for high predation value i.e. $p = 0.35$ the fluctuations are evident at lower value of harvesting, i.e. $H_x = 0.43$ (see Fig.~\ref{fig7}B). This indicates that the risk of extinction of prey fish increases with the predation level and the prey harvesting rate.

Next, we analyze the dynamics for a range of predation rates and harvesting rate of prey and predator fish, which are depicted in two-parameter bifurcation diagrams (Figs.~\ref{fig7}C and \ref{fig7}D). We find interesting dynamics in a wide range of parameter values. For higher predation rate $p$ and lower values of prey harvesting rate $H_x$, we observe saddle-node bifurcation (see Fig.~\ref{fig7}C). With an increase in $H_x$ and lower $p$ value, near $0.29$, we get Hopf bifurcation. One may note that at lower $p$ value, the system experiences monostable state. Further, from Fig.~\ref{fig7}D, we find that a high predator harvesting rate $H_y$ and high $H_x$ exhibits cyclic dynamics. However, for low values of $H_y$, cyclic dynamics no longer exist and exhibits saddle-node bifurcation.

\subsection{Influence of social norms}

Due to over-harvesting, fisheries are prone to sudden collapse, which is evident from our previous analysis (see Fig.~\ref{Fig1} and Fig.~\ref{Fig5}). Therefore, for the future sustainability of fisheries, fishing management strategies are required to strengthen such systems' resilience. Here, we incorporate social norms in the two species model \eqref{Eqn1}. We apply the concept of opinion dynamics to the prey harvesting rate $H_x$, which now takes the form $H_x(1-s)$. Here, $s$ is the proportion of the population that adopts the opinion of conservation. The opinion dynamics is described by the concept of imitation dynamics of evolutionary game theory \cite{lade2013regime}. It captures human behavior, especially how humans like to imitate successful strategies \citep{hofbauer1998evolutionary}. Here, we consider the opinions of either "adapting conservation" or "not adapting conservation", and each one is associated with a payoff. Each individual can sample another individual with a sampling rate. In particular, if the sampled person has other opinions with a higher payoff, the individual switches to the sampled person's opinion with the expected gain in the payoff. The following is the resulting coupled socio-ecological model with social norms:
\begin{subequations}\label{Eqn10}
\begin{align}
 \frac{dx}{dt} &= I+ r_x x (1-\frac{x}{K_x})- \frac{p y x^2}{h^2+x^2}-H_x (1-s) x,\\
 \frac{dy}{dt} &= r_y y (1-\frac{y}{K_y})+\frac{c p y x^2}{h^2+x^2}-H_y y,\\
 \frac{ds}{dt} &= k s (1-s) \left(d(2s-1)-w+\frac{1}{x+c_1} \right),
\end{align}
\end{subequations}
where $k$ and $d$ are the sampling rate and strength of the injunctive social norms, respectively.  $w$ is the total cost of conservation, and $c_1$ is a control parameter that controls the prey fish density with the payoff of conserving fish. In particular, for a small value of $x$, $\frac{1}{x+c_1}$ motivates fish conservation, as the utility to conserve prey fish increases when fish densities become very low. All the other parameters are the same as in \eqref{Eqn1}.
\begin{figure}
\begin{center}
\includegraphics[width=0.85\columnwidth]{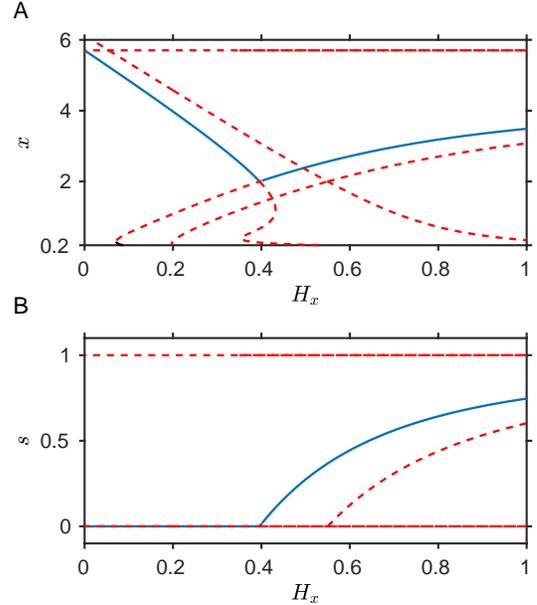}
\caption{\label{fig8} (A-B) Codimension-one bifurcation diagrams for the eqn~\ref{Eqn10}. (A) One parameter bifurcation diagram of the prey fish ($x$) at different prey harvesting $(H_x)$. (B) One parameter bifurcation diagram of the social norms at different prey harvesting $(H_x)$. Blue lines and red dotted lines are denoting stable attractor and unstable attractor respectively. Other parameters are $I=0.1$, $c=0.2$, $H_y=0.15$, $r_x=0.75$, $K_x=6$, $h=0.2$, $r_y=0.25$, $K_y=4$, $p=0.15$, $k=0.2$, $w=0.3$, $c_1=0.5$, $d=0.1$.}
\end{center}
\end{figure}


Here, we examine the qualitative dynamics of the model \eqref{Eqn10}, using one parameter bifurcation diagrams by varying $H_x$ (see Fig.~\ref{fig8}). The coupled model exhibits transcritical and saddle-node bifurcations (see Fig.~\ref{fig8}A). We can say that over-harvesting is a significant cause of the collapse of fish dynamics (see Fig.~\ref{Fig1}). However, a very different outcome occurs when small conservation cost ($w=0.3$) and weaker social norms ($d=0.1$) is applied. In the vicinity of the tipping point, where $H_x$ takes the value around $0.4$ (see Fig.~\ref{fig8}A), we observe that on adapting conservation opinion, i.e., considering $s>0$ (see Fig.~\ref{fig8}B), the collapse of fish could be prevented. It is possible to maintain moderate fish densities despite higher harvesting rates, as clearly depicted in Fig.~\ref{fig8}A.

\section{Discussion \label{dis}}

Increasing anthropogenic pressure, in the form of increasing population density \cite{barrett2020social}, invoke profound global change that has many future repercussions, specifically increasing harvesting rates, which may flip some lakes or oceans from a stable state to an uncertain condition \citep{roberts1999extinction, sadovy2001threat, fryxell2017supply}.  In this paper, we investigate the response of an ecological model towards harvesting pressures, under the influence of demographic and environmental stochasticity. We proceed by analyzing the behavior of a two-species fish model \eqref{Eqn1}. We find that the model involves a hysteresis loop (see Fig.~\ref{Fig1}), which primarily instincts the possibility of sudden critical transitions. Here we have used the master equation, which follows the Markov process, to study the effects of demographic stochasticity on the model. For different parameter values, stochastic potentials are calculated to examine the influence of demographic noise on the resilience of steady states whose persistence is supported by MFPT.  MFPT for the considered model reveals that the upper steady state is more resilient than the lower steady state. Further, for environmental stochasticity the system experiences more fluctuations when compared to demographic stochasticity. In the case of environmental noise, we calculate the system trajectory by applying the Euler-Maruyama method. Our results show that in the presence of demographic and environmental stochasticity, an increase in the harvesting rate of prey species induces a critical transition from a high fish density state to a low fish density state.

Moreover, we calculate a few generic early warning signals to forewarn the chance of critical transitions. Population extinction has always been a major concern, but anticipating species extinction remained challenging \citep{ludwig1999meaningful}. This lead to a quest for robust indicators of critical transitions in ecology \citep{dakos2008slowing, drake2010early}. We found that the indicator AR(1) could successfully forewarn an impending transition for environmental stochasticity but fails to provide any warning in the case of demographic noise. However, variance and conditional heteroskedasticity work well in predicting transitions for both types of stochasticity. Further, we also emphasized the importance of considering management steps for the better future of the fisheries. The extended socio-ecological model gives the qualitative dynamics of prey fish densities and social norms and their relationship with harvesting ($H_x$). We observe that the application of social norms leads to the growth of prey species, which otherwise would have resulted in an inevitable collapse of the system.

The anticipation of critical transitions could prevent sudden catastrophes and thus enhance the stability of a system. Furthermore, using various early warning indicators that could predict an upcoming shift would be invaluable. For instance, predicting regime shifts in marine ecosystems using EWS could impede the switch to low fish density and result in the persistence of the marine population. However, theories suggest that it is necessary to analyze specific situations under which generic early warning indicators fail to work effectively. Our results firmly outline the need to develop robust early warning indicators with a good understanding of which of them might be most convenient in terms of perceptibility and reliability.

On a final note, our work illustrates the influence of natural and anthropogenic factors on an ecological system. Further, understanding critical transitions in higher-dimensional systems, as in network models, would be a useful research area. Moreover, developing models to explore rate-induced tipping and considering the effects of different noise types on the predictability of early warning signals is an important future direction. Nevertheless, predicting the distance to critical transitions remains an important research question, together with predicting the system's state after these sudden transitions.

\begin{acknowledgments}
 S.S. acknowledges the financial support from DST, India under the scheme DST-Inspire [Grant No.: IF160459].  P.S.D. acknowledges financial support from the Science \& Engineering Research Board (SERB), Govt. of India [Grant No.: CRG/2019/002402].
\end{acknowledgments}



%

\end{document}